\documentclass[final]{svjour2}
\usepackage{graphicx}
\usepackage{rotating}
\usepackage{amssymb}
\usepackage{mathptmx}
\usepackage[numbers]{natbib}
\makeatletter
\journalname{Journal of Low Temperature Physics}

\def\6#1{{\underline{#1}}}
\def\m6#1{{\underline{#1}\,}}

\newdimen\Tdim
\def\ispan{{\setbox0=\hbox{i}%
\Tdim\ht0\advance\Tdim\dp0\rule[-\dp0]{0pt}{\Tdim}}}
\def\jspan{{\setbox0=\hbox{j}%
\Tdim\ht0\advance\Tdim\dp0\rule[-\dp0]{0pt}{\Tdim}}}
\def\Tspan#1{{\setbox0=\hbox{#1}%
\Tdim\ht0\advance\Tdim\dp0\advance\Tdim.55ex\rule[-\dp0]{0pt}{\Tdim}\box0}}

\def\be{\begin{eqnarray}}
\def\ben{\begin{eqnarray*}}
\def\ee{\end{eqnarray}}
\def\een{\end{eqnarray*}}

\def\=:{=\hspace{-.7em}\raisebox{1.1ex}{.}\hspace{.1em}\raisebox{-0.2ex}{.} }

\newcommand {\1}[1]{\frac{1}{#1}}

\newcommand {\beq}{\begin{eqnarray}}
\newcommand {\eeq}{\end{eqnarray}}
\newcommand {\non}{\nonumber\\}

\bibpunct{}{}{,}{s}{}{,}

\begin{document}

\newcommand{\hdblarrow}{H\makebox[0.9ex][l]{$\downdownarrows$}-}
\title{Vortex molecules in Bose-Einstein condensates}

\author{Muneto Nitta$^1$ \and Minoru Eto$^2$ \and Mattia Cipriani$^3$}

\institute{1:Department of Physics, and Research and Education Center for Natural Sciences, Keio University, Hiyoshi 4-1-1, Yokohama, Kanagawa 223-8521, Japan\\
\email{nitta@phys-h.keio.ac.jp}
\\2: Department of Physics, Yamagata University, 
Yamagata 990-8560, Japan
\\ 3: University of Pisa, Department of Physics ``E. Fermi'', INFN, Largo Bruno Pontecorvo 7, 56127, Italy}

\date{XX.XX.2013}

\maketitle


\begin{abstract}

Stable vortex dimers are known to exist in coherently coupled two component  Bose-Einstein condensates (BECs).
We construct stable vortex trimers 
in three component BECs 
and find that 
the shape can be controlled by changing 
 the internal coherent (Rabi) couplings.
Stable vortex $N$-omers are also constructed in 
coherently coupled $N$-component BECs.
We classify all possible $N$-omers in terms of the mathematical graph theory. 
Next, we study effects of the Rabi coupling in vortex lattices in two-component BECs. We find how the vortex lattices without the Rabi coupling known before
are connected to the Abrikosov lattice of integer vortices with increasing the Rabi coupling.
In this process, 
vortex dimers change their partners in various ways 
at large couplings.
We then find that 
the Abrikosov lattices  are robust in three-component BECs.

\end{abstract}

\section{Introduction}

The study of Bose-Einstein condensates (BECs) can provide an ideal opportunity to examine the dynamics of exotic vortices because these vortices can theoretically be explained via 
a quantitative description using the mean field theory 
by the Gross-Pitaevski (GP) equation
and further, the vortices can be experimentally controlled to a large degree. 
Vortices in multi-component BECs have been realized experimentally \cite{Matthews,Schweikhard:2004}, 
and the structures of vortices in spinor BECs
\cite{Ho:1998,Semenoff:2006vv,Turner:2009}
and in mixtures of multiple species 
and/or multiple hyperfine states of the same atom  
\cite{Mueller:2002,Kasamatsu:2003,Kasamatsu:2004,
Kasamatsu:2005,Kasamatsu:2009,Eto:2011wp,
Aftalion:2011,Aftalion:2012,Kuopanportti:2012,
Eto:2012rc,Eto:2013spa,Cipriani:2013nya,Cipriani:2013wia}
are considerably  
richer than those formed of single components;  
one fascinating feature in such cases is that  
vortices are fractionally quantized, 
and the other is that sets of fractional vortices constitute 
vortex molecules.   
In the case of multiple hyperfine spin states of BECs, 
internal coherent (Rabi) couplings between multiple components 
can be introduced by Rabi oscillations, 
similar to Josephson couplings in 
multi-gap superconductors. 
As in two-gap superconductors \cite{Tanaka:2001}, 
a sine-Gordon domain wall of the phase difference of two components 
(phase soliton) appears \cite{Son:2001td}. 
In this light, the advantages of BECs are that Rabi coupling can be tuned experimentally 
and that in contrast to two-gap superconductors \cite{Babaev:2002,Goryo:2007}, 
a vortex dimer, 
thas is two fractional vortices connected by a phase soliton, 
can exist stably because a repulsion between 
two fractional vortices \cite{Eto:2011wp} 
is sufficiently strong  
in the absence of the Meissner effect,
to be balanced with the tension of 
a sine-Gordon domain wall connecting them 
\cite{Kasamatsu:2004}.

Recently, we have observed stable-vortex trimers 
in three-component BECs, and 
we have shown that the shape and the size of the molecule 
can be controlled by varying the strength of  
the Rabi couplings \cite{Eto:2012rc}. 
We have further constructed stable-vortex $N$-omers, 
that is, molecules made of $N$ fractional vortices in 
$N$-component BECs,
where each condensate wave function 
has a nontrivial winding number around one of 
the $N$ fractional vortices \cite{Eto:2013spa}.
$N$-omers with $N \ge 3$ 
exhibit several novel properties that dimers do not possess, that is, 
the existence of chirality pairs 
and metastable states such as twist, holding, and 
capture, which are properties exhibited by 
chemical molecules.
The mathematical graph theory is useful to
classify the vortex $N$-omers 
whose number exponentially increases as $N$ increases. 
All possible graphs have been constructed 
for $N=3,4,5$ by imaginary time propagation \cite{Eto:2013spa} 
with the phase winding and a constant density 
fixed at the boundaries, thereby our results
imply that the vortex $N$-omers are formed in rotating BECs. 

The Rabi coupling in vortex lattices
in two-component BECs under rotation 
has been studied systematically \cite{Cipriani:2013nya}.
The triangular and square vortex lattices 
are connected to
the Abrikosov lattices of integer vortices at the strong Rabi coupling, 
while in the intermediate Rabi couplings,  
we find that vortices change their partners 
in various ways depending on the inter-component coupling. 
Vortex lattices in a three-component BEC 
with three kinds of fractional vortices winding one of three components 
have been studied recently \cite{Cipriani:2013wia}.  
We always find triangular ordered 
vortex lattices, where 
three kind of fractional vortices are placed in 
order without defects.

\section{$N$-component BECs}

We consider coherently coupled 
miscible 
$N$-component BECs of atoms with mass $m$, 
described by
the condensate wave functions $\Psi_i$ ($i=1,2,\cdots,N$) with the GP energy functional
\beq
E &=& \sum_{i,j=1}^N\int d^2x 
 \bigg(\frac{\hbar^2}{2m} 
\left| \left(-i \nabla - {m\over \hbar} {\Omega} \times {\bf r} \right)\Psi_i \right|^2 \delta_{ij}
+ \frac{g_{ij}}{2}|\Psi_i|^2 |\Psi_j|^2 \non
&& + (V_{\rm eff.}- \mu_i|\Psi_i|^2) \delta_{ij} - \omega_{ij}\Psi_i^*\Psi_j
\bigg),
\label{eq:gp}
\eeq
where atom-atom interactions are characterized by 
the coupling constants $g_{ij} = 4\pi \hbar^2 a_{ij}/m$ 
with s-wave scattering lengths $a_{ij}$, 
$\mu_i$ denotes the chemical potential, and 
$\omega_{ij} = \omega_{ji}$ ($\omega_{ii} = 0$)
denotes the internal coherent coupling due to 
Rabi oscillations 
between the $i$-th and $j$-th components.
$\Omega$ is the rotation frequency of the system, and 
$V_{\rm eff.} = V_{\rm ex.} - m\Omega^2 r^2/2$ 
is the effective trapping potential of the external potential $V_{\rm ex}$ 
combined with the centrifugal potential.
We mainly consider the case where 
$g_{ii} \equiv g$, 
 and $\mu_i \equiv \mu$,
and $g_{ij} \equiv \tilde g$ ($i\neq j$). 
It is straightforward to consider the general case but
all the results below are essentially unchanged from our simplest choice. 
We mostly consider the miscible case $g>\tilde g$.  
The symmetry of the Hamiltonian depends on the coupling constants: 
\beq
 \left\{
 \begin{array}{c}
   U(N) , \quad g = \tilde g , \quad \omega_{ij} = 0, \cr 
   U(1)^N , \quad g \neq \tilde g , \quad \omega_{ij}=0 ,\cr
   U(1) ,\quad g \neq \tilde g , \quad \omega_{ij}\neq 0
 \end{array} 
\right.  \label{eq:symmetry}
\eeq

Let us consider an infinite uniform system 
with $\Omega=0$ and $V_{\rm eff.}=0$. 
When all the internal coherent couplings are equal {\it i.e.},
$\omega_{ij}=\omega$, 
the condensations of the ground state are 
\beq 
 |\Psi_i| = v \equiv \sqrt{\frac{\mu + (N-1)\omega}  
 {g+(N-1)\tilde g}},
\eeq
with $i=1,2,\cdots,N$. 
These amplitudes are modified  
in the case $\omega_{ij} \neq \omega_{i'j'}$, and
one should solve the variational problem $\delta E/ \delta \Psi_i = 0$ to 
determine $v_i= |\Psi_i|$.

As long as the internal coherent couplings $|\omega_{ij}|$ are maintained sufficiently small with respect to the other couplings
(we choose $g \sim \tilde g = {\cal O}(10^3)$, $\mu = {\cal O}(10^2)$, and $\omega = {\cal O}(10^{-2})$),
the symmetries (\ref{eq:symmetry})
are nearly intact. 
These symmetries are spontaneously broken 
in the ground state. 
Consequently, vortices that are 
quite different for $g = \tilde g$ and $g > \tilde g$ can appear.

The internal coherent couplings 
$- \omega_{ij}\Psi_i^*\Psi_j = 
- 2v_iv_j\omega_{ij}\cos(\theta_i - \theta_j)$,
with $\theta_i = \arg \Psi_i$, 
give gaps to the Legette modes $\theta_i - \theta_j$. 
If we approximate the amplitudes of 
the condensate wave functions to be constant, 
$|\Psi_i| = v_i e^{i \theta_i}$,
the truncated energy functional is obtained as
\beq 
{\cal E}_{\rm phase} = \sum_{i=1}^N\frac{\hbar^2}{2m}(\vec \nabla \theta_i)^2 
- 2\sum_{i>j}^N\omega_{ij}v_iv_j\cos (\theta_i-\theta_j).
  \label{eq:reduced-energy}
\eeq
For $\omega_{ij} > 0$,  
all the phases $\theta_i$ and $\theta_j$ coincide, {\it i.e.}, $\theta_i = \theta_j$ 
in the ground state. 
This model allows sine-Gordon domain walls (phase solitons).
For two components $N=2$, 
the conventional sine-Gordon domain walls 
are allowed \cite{Son:2001td}. 
The same phase solitons were also predicted 
in two-gap superconductors \cite{Tanaka:2001}.
For higher $N$, domain wall junctions may exist,
but no analytic solutions are available.

\section{Vortex molecules}

In this section, we discuss one integer vortex which is 
split into fractional vortices. 
For this purpose, it is enough to consider 
an infinite uniform system with $\Omega=0$ and $V_{\rm eff.}=0$ 
with the boundary condition for the phase winding.

\subsection{Fractional vortices}
First, 
let us consider the case that 
all $\omega_{ij}$'s are zero. 
A vortex winding around only the $i$-th component 
($i=1,\cdots,N$), 
$\Psi_i = v_i \rho(r) e^{i\theta}$,  
in polar coordinates $r$ and $\theta$,   
with the other components almost constant, 
is fractionally quantized as
\beq
 \oint d {\bf r} \cdot {\bf v}_{\rm s} = {{v_i^2}\over {\sum_i} v_i^2} {h \over m},
\eeq
with the superfluid velocity ${\bf v}_{\rm s}$,
and it is stable. 
We label it by an $N$-vector 
$(0,\cdots,1,\cdots,0)$ where $i$-th component is $1$. 
The interaction between two vortices 
in the same component separated by the distance $R$ 
is $F \sim 1/R$ as
the same as 
that between two integer vortices in the scalar BEC,
while the interaction between vortices in different components, say $(1,0)$ and $(0,1)$, 
is $F \sim g_{12} (\log R/\xi - 1/2)/R^3$, 
which is repulsive in our case of $g_{12}>0$ \cite{Eto:2011wp}.   



For $g=\tilde g$, an axisymmetric giant vortex appears. 
This can be interpreted as 
a $\mathbb{C}P^{N-1}$ skyrmion \cite{Eto:2013spa}.  
On the other hand, 
for the miscible case with $g > \tilde g$, 
there appear $N$ fractional vortices associated with
the broken $U(1)^N$ symmetry,
which are connected by domain walls, thereby
resulting in molecules with $N$ vortices.
The reduced energy functional in Eq.~(\ref{eq:reduced-energy}) shows that $N-1$ domain walls are attached 
to the $i$-th vortex in the direction $\theta_i = \theta_j$. 
Therefore, this vortex cannot remain stable in an isolated state, and it must be connected to the boundary or to vortices winding around the other components forming the molecules.

\subsection{Vortex dimers} 
\begin{figure}[h] 
\vspace{-0.5cm}
\begin{center}
\includegraphics[width=4cm,clip]
{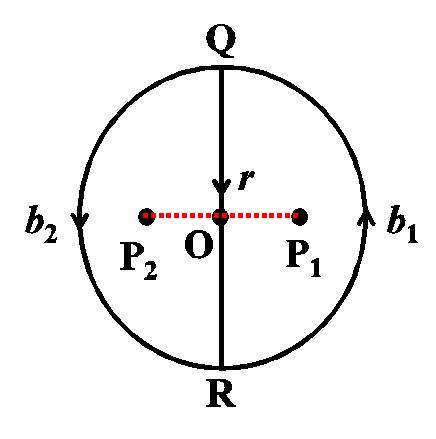}
\caption{
The $(1,0)$, and $(0,1)$ vortices 
are placed at P$_1$, P$_2$, respectively. 
$b_i$ ($i=1,2$) correspond to 1/2 circles at the boundary 
at spatial infinities,   
and $r$ is the path connecting Q and R 
passing through the origin O. 
The $(1,0)$ and $(0,1)$ vortices are 
encircled by $b_1 + r$ and  $b_2 -r$, 
respectively.
\label{fig:dimer}}
\end{center}
\end{figure}

Vortex dimers are molecules of two vortices appearing in
two-component BECs \cite{Kasamatsu:2004}. 
Vortices winding around the first and the second components denoted by $(1,0)$ and $(0,1)$, respectively, 
are connected by a sine-Gordon domain wall 
\cite{Son:2001td}. 
To see this, we note that 
the reduced energy functional is
\beq 
{\cal E}_{\rm phase} 
= \frac{\hbar^2}{4m}[\vec \nabla (\theta_1 + \theta_2)]^2 
+ \frac{\hbar^2}{4m}[\vec \nabla (\theta_1 - \theta_2)]^2 
- 2 \omega_{12}v_1v_2\cos (\theta_1-\theta_2).
  \label{eq:reduced-energy-2comp}
\eeq
We see that the overall phase $\theta_1 + \theta_2$ decouples 
and the phase difference $\theta_1 - \theta_2$ 
becomes the sine-Gordon model. 
Let us place the $(1,0)$ and $(0,1)$ vortices 
at P$_1$ and $P_2$, respectively in Fig.~\ref{fig:dimer}.
The $(1,0)$ and $(0,1)$ vortices 
are enclosed by the paths $b_1 + r$ and $b_2 - r$, 
respectively. 
Interpreting the label $(a,b)$ as a generator 
acting as $(\Psi_1,\Psi_2) \to (\Psi_1 e^{ia \alpha}, \Psi_2 e^{ib \alpha})$,
we decompose the generators $(1,0)$ and $(0,1)$ of vortices as
\beq
 (1,0) = \1{2} (1,1) + \1{2} (1,-1) , \quad
 (0,1) = \1{2} (1,1) - \1{2} (1,-1).
\eeq
Since the total configuration has unit winding 
for the overall phase (gauge), 
the paths $b_1$ and $b_2$ correspond to 
the half windings generated by $(1,1)$.
Then, the other generator $(1,-1)$ acts 
along the path $r$:
\beq 
 \Psi_1 (x=0,y) = e^{i \pi g(y)} |\Psi_1|, \quad 
 \Psi_2 (x=0,y) = e^{-i \pi g(y)} |\Psi_2|, 
\eeq
with monotonically increasing function 
$g(y)$ with the boundary conditions
$g(y \to -\infty) = 0$ and 
$g(y \to -\infty) = 1$.  
We find that the phase difference
$\theta_1 - \theta_2$ changes by $2\pi$ 
along the path $-r$  
and that a sine-Gordon domain wall exists
on the path due to the potential term  
$V = 2 \omega_{12}v_1v_2\cos (\theta_1-\theta_2)$
in (\ref{eq:reduced-energy-2comp}).
This domain wall connects the two 
fractional vortices. 

In order to have a stable configuration, we need 
$g_{12}>0$ so that two vortices repel each other 
for $\omega_{12}=0$. 
For $\omega_{12} \neq 0$, the domain wall tension 
can be balanced with the repulsion between them, 
to yield a stable configuration. 
Numerical solutions of stable 
vortex dimers were obtained in Ref.~\cite{Kasamatsu:2004}.

Vortex dimers were also proposed in two-gap superconductors 
\cite{Babaev:2002}. 
However, in this case, the gauge symmetry is local.  
Consequently, the repulsion between 
$(1,0)$ and $(0,1)$ vortices are exponentially 
suppressed by the Meissner effect  
and cannot be balanced with 
a constant force of the domain wall tension 
(confinement).  
Instead, the deconfinement mechanism similar to 
the Berezinskii-Kosterlitz-Thouless transition at finite temperature 
was proposed \cite{Goryo:2007}.

\subsection{Vortex trimers}
Three-component BECs admit 
vortex trimers, 
molecules of three vortices  
labeled as $(1,0,0),(0,1,0)$, and $(0,0,1)$,  
winding around the first,  second, and third component by $2\pi$, 
respectively. 
The energy of each vortex is logarithmically divergent 
when $\omega_{ij}=0$ and linearly divergent when $\omega_{ij}\neq 0$, 
if the system size is infinite.

\begin{figure}
\vspace{-0.5cm}
\begin{center}
\begin{tabular}{ccc}
\includegraphics[width=4cm,clip]
{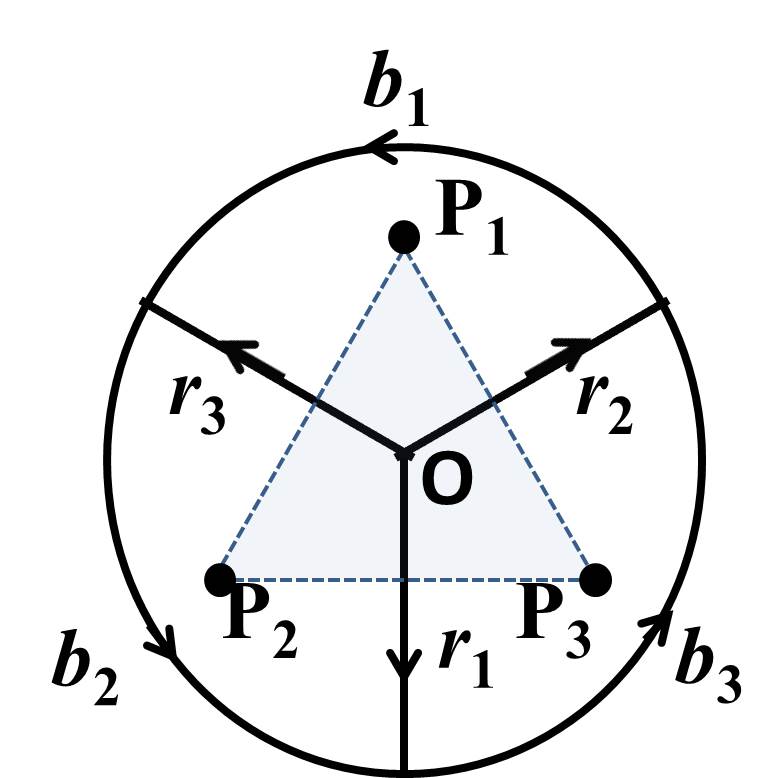} &
\includegraphics[width=4cm]{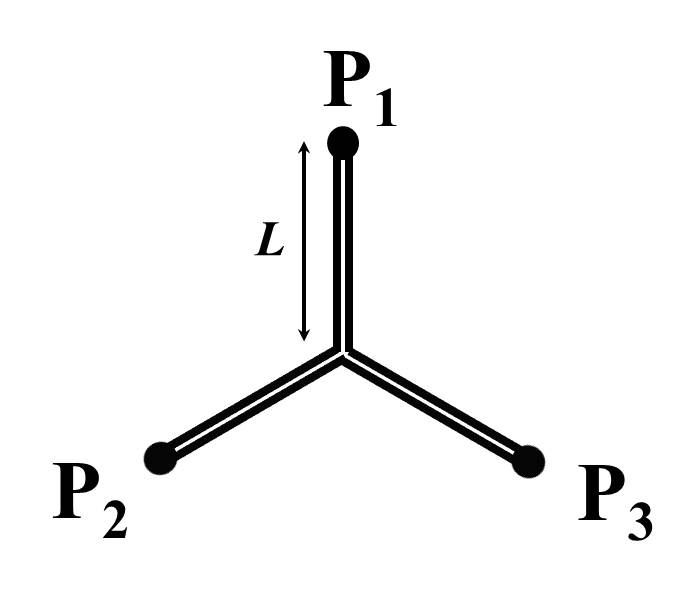} &
\includegraphics[width=1.8cm]{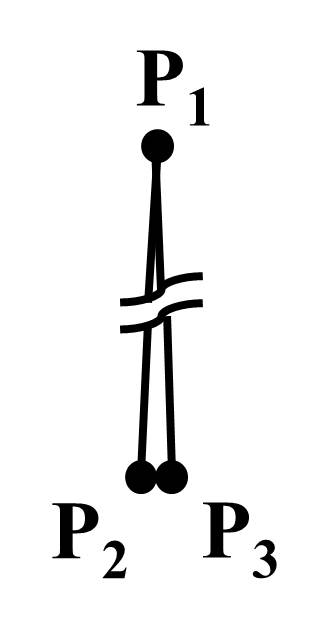}  \\
(a) & (b) & (c)
\end{tabular}
\caption{
(a) 
The $(1,0,0),(0,1,0)$, and $(0,0,1)$ vortices 
are placed at P$_1$, P$_2$ and P$_3$, respectively. 
$b_i$ ($i=1,2,3$) corresponds to 1/3 circles at the boundary,   
and $r_i$ corresponds to the radial paths from the origin O 
to the circle at the boundary. 
The $(1,0,0)$, $(0,1,0)$, and $(0,0,1)$ vortices are 
encircled by $b_1 -r_3 + r_2$,  $b_2 -r_1 + r_3$, and 
$b_3 -r_2 + r_1$, respectively.
(b) A domain wall junction.
(c) The two vortices P$_2$ $(0,1,0)$ 
and P$_3$ $(0,0,1)$ together are placed 
at the same position very far from 
the vortex P$_1$ $(1,0,0)$. 
They are connected by a sine-Gordon domain wall. 
\label{fig:molecule}}
\end{center}
\end{figure}
\begin{figure}
\begin{center}
\includegraphics[width=5cm]{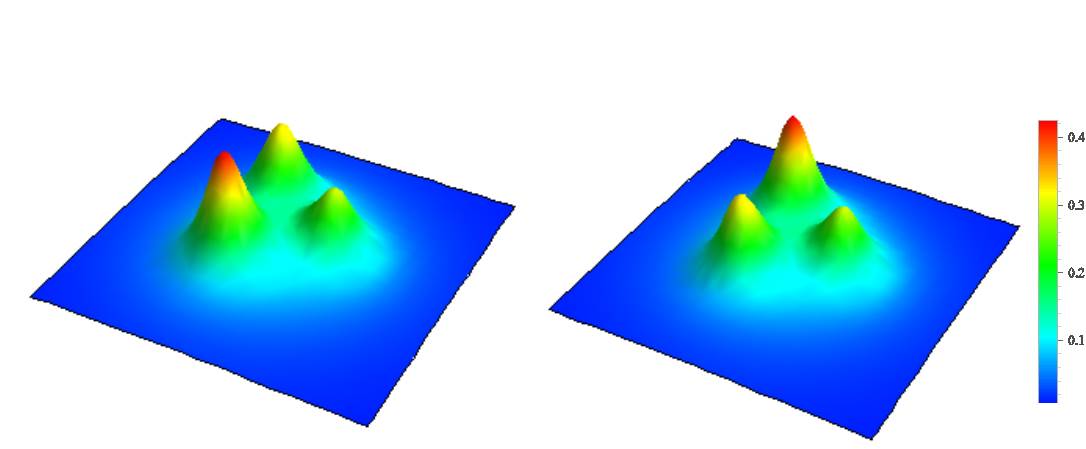}
\caption{Two trimers in a three-component BEC ($g_{11} \neq g_{22} \neq g_{33} \neq g_{11}$).
The left panel is a mirror image of the right panel.}
\label{fig:chiral}
\end{center}
\end{figure}

Following Ref.~\cite{Nitta:2010yf}, 
let us consider that the $(1,0,0),(0,1,0)$, and $(0,0,1)$ vortices 
are placed at the edges 
(P$_1$, P$_2$, and P$_3$, respectively) of 
a triangle, as shown in Fig.~\ref{fig:molecule}(a).
In the large circle, the total configuration is 
the integer vortex $(1,1,1)$.
In other words, the integer vortex has 
an internal structure made of 
three fractional vortices.
Instead of the $U(1)^3$ generators $(1,0,0),(0,1,0)$, and $(0,0,1)$, 
let us prepare four linearly dependent generators: 
the gauge rotation $(1,1,1)$ and 
three gauge-invariant rotations $(0,-1,1)$, $(1,0,-1)$, and $(-1,1,0)$. 
In these generators, the winding of the $(1,0,0),(0,1,0)$, 
and $(0,0,1)$ vortices can be decomposed into
\beq
&&{\rm P}_1:\quad 
 (1,0,0)=\1{3}(1,1,1)    +0(0,-1,1)+\1{3}(1,0,-1)-\1{3}(-1,1,0),\non
&&{\rm P}_2:\quad 
 (0,1,0)=\1{3}(1,1,1)-\1{3}(0,-1,1)    +0(1,0,-1)+\1{3}(-1,1,0),\non
&&{\rm P}_3:\quad 
 (0,0,1)=\1{3}(1,1,1)+\1{3}(0,-1,1)-\1{3}(1,0,-1)+    0(-1,1,0).
\eeq
We see that all $b_i$ ($i=1,2,3$) correspond to $2\pi/3$ rotation of 
the gauge generator $(1,1,1)$, 
and consequently that 
these vortices have 1/3 quantized 
circulations $h/3m$
for $v_1=v_2=v_3$.  
The $(1,0,0)$, $(0,1,0)$, and $(0,0,1)$ vortices are 
encircled by $b_1 -r_3 + r_2$,  $b_2 -r_1 + r_3$, and 
$b_3 -r_2 + r_1$, respectively (Fig.~\ref{fig:molecule}). 
Therefore, we can identify the paths 
$\pm r_1$, $\pm r_2$ and $\pm r_3$ 
corresponding to $\pm 2\pi/3$ of  
the phase rotations by $(0,-1,1),(1,0,-1)$, and $(-1,1,0)$, 
respectively.
The phase rotation along the radial path $r_i$ 
can be written up to constant phases as 
\beq
&& r_1: \quad 
  \Psi_1 = |\Psi_1|,\quad 
  \Psi_2 = e^{-(2\pi i/3) f(r)}|\Psi_2|, \quad 
  \Psi_3 = e^{(2\pi i/3) f(r)}|\Psi_3| ,\non
&& r_2: \quad 
  \Psi_1 = e^{(2\pi i/3) f(r)}|\Psi_1| ,\quad 
  \Psi_2 = |\Psi_2|,\quad 
  \Psi_3 = e^{-(2\pi i/3) f(r)}|\Psi_3| ,\non
&& r_3: \quad 
  \Psi_1 = e^{-(2\pi i/3) f(r)}|\Psi_1| ,\quad 
  \Psi_2 = e^{(2\pi i/3) f(r)}|\Psi_2| ,\quad 
  \Psi_3 = |\Psi_3| ,\label{eq:r-pathes}
\eeq
where a function $f(r)$ has the boundary conditions 
$f(r=0)=1$ and $f(r\to \infty)=0$. 
Since the potential term does not vanish 
at the origin O, the domain walls cannot connect straight 
two among P$_1$, P$_2$ and P$_3$.
We then expect that the three domain walls are 
bent to constitute a domain wall junction 
as in Fig.~\ref{fig:molecule}(b). 
Each leg of the junction is a sine-Gordon 
domain wall. 
This can be justified if we separate one of 
vortices, say P$_1$, far from the others 
as in Fig.~\ref{fig:molecule}(c). 
In the limit that P$_1$ is separated infinitely, 
we have $\theta_2 = \theta_3$, 
and  
the truncated energy functional 
(\ref{eq:reduced-energy}) for $N=3$ 
is reduced to 
${\cal E}_{\rm phase} 
= \sum_{i=1}^3\frac{\hbar^2}{2m}(\vec \nabla \theta_i)^2 
- 2 v_1(\omega_{12}v_2+\omega_{13}v_3)\cos (\theta_1-\theta_2)
$.

Vortex trimers exhibit two new properties that the dimers do not have.
The first reported in Ref.~\cite{Eto:2012rc}  
is that the shape of the triangle can be changed
by tuning the parameters in Eq.~(\ref{eq:gp}).
The second is the chirality of the triangle. 
This can be easily seen 
when the coupling constants are generic, 
so that the vortices have different sizes,  
as seen in Fig.~\ref{fig:chiral}.
As shown in Fig.~\ref{fig:chiral}, 
the left and right configurations 
cannot be transformed to each other by 
a rotation 
but can be transformed by a mirror.
They are energetically completely degenerate.

\subsection{Vortex tetramers and $N$-omers} 

\begin{figure}
\begin{center}
\includegraphics[width=11cm]{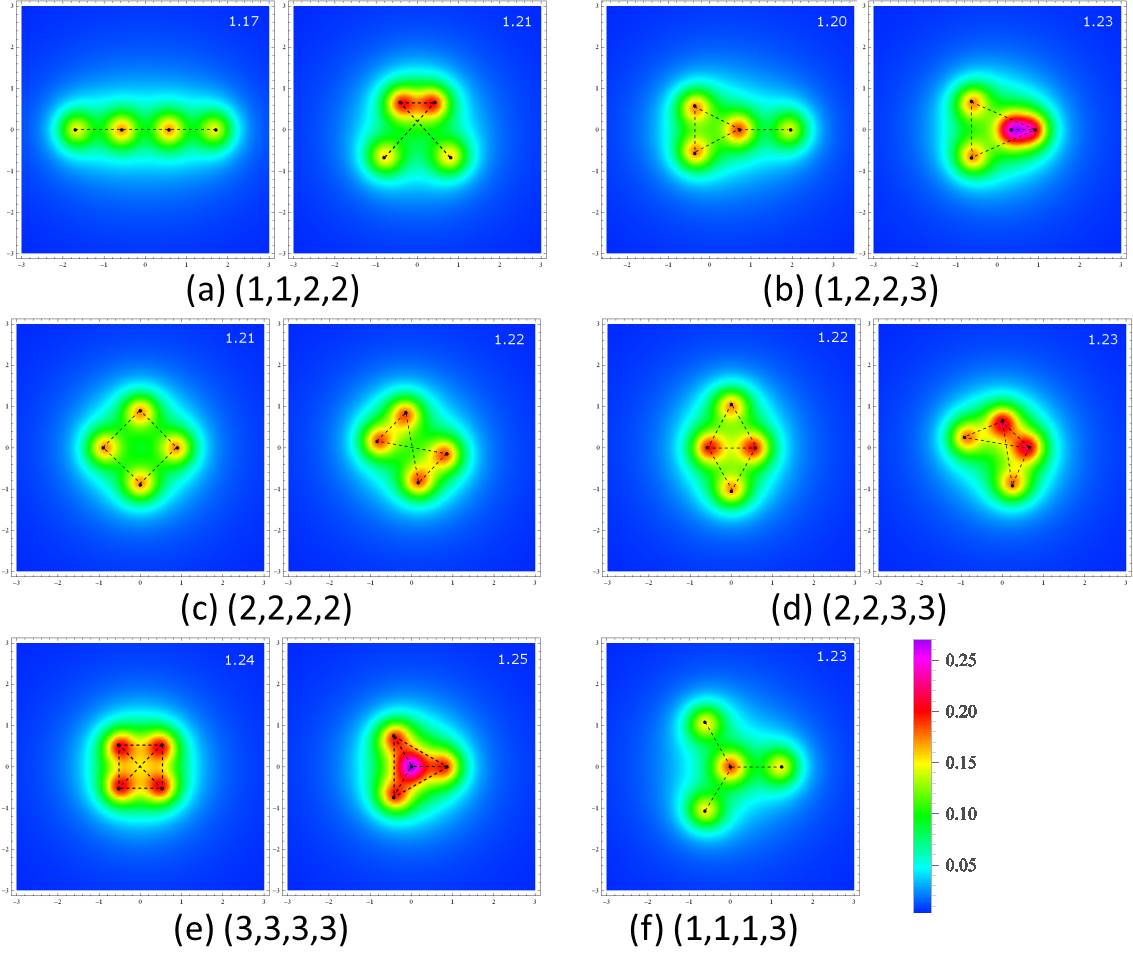}
\caption{All possible connected tetramers in a four-component BEC. 
We set 
$m=\hbar=1$, $g = 10^3$, $\tilde g=900$, 
and $\mu = 10^2$. 
We also set
$\omega_{ij}=0.05$ and $\omega_{ij} = 0$ for
the connected and disconnected pairs, respectively. 
The dots and dotted lines indicate the vortex center of each component and 
the Rabi couplings between components, respectively. 
The color maps represent the energy density.  
The number at the upper-right corner in each subfigure 
represents the total energy. 
For (a), (b), (c), (d) and (e),  we have found 
the stable configurations (the left) and 
the metastable configurations (the right). 
\label{fig:4comp}
}
\end{center}
	\begin{minipage}{0.50\textwidth}
		\includegraphics[scale=0.60]{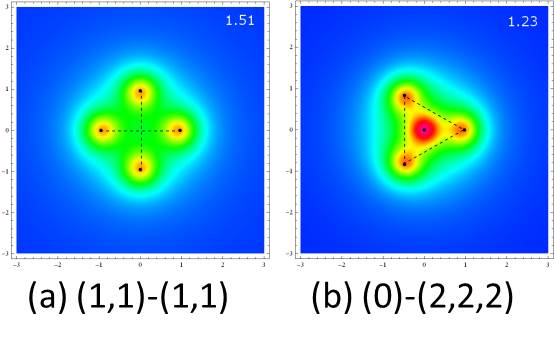}
	\end{minipage}\quad
	\hfill
	\begin{minipage}{0.50\textwidth}
		\vspace{-0.5cm} 
\includegraphics[scale=0.5]{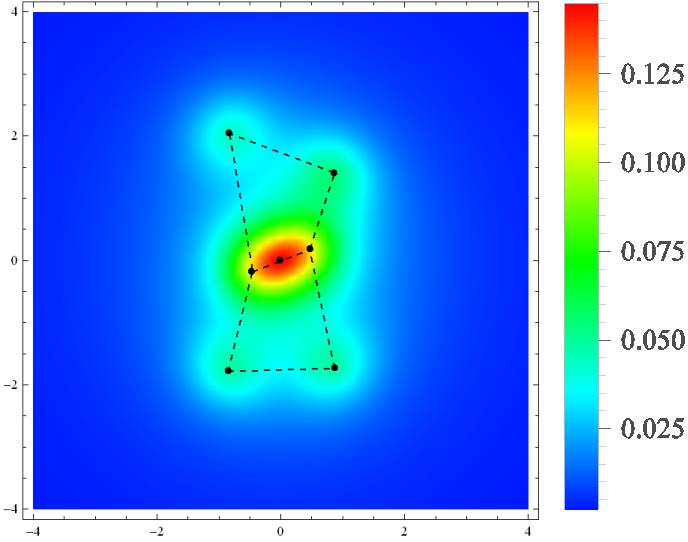}
	\end{minipage}
	\begin{minipage}[t]{0.50\textwidth}
		\caption{Disconnected tetramers in a four-component BEC. 
The parameters are the same with Fig.~\ref{fig:4comp}. 
\label{fig:4comp2}
}
	\end{minipage}\quad
	\hfill
	\begin{minipage}[t]{0.50\textwidth}
		\caption{A vortex heptamer designed to be the orion 
in a seven component BEC.
We set $(\omega_{12},\omega_{23},\omega_{34},\omega_{45},\omega_{16},\omega_{56},\omega_{37},\omega_{67})=
(0.02,0.001,0.02,0.02,0.05,0.02,0.2,0.2)$ and the rests to be zero.
\label{fig:orion}}
	\end{minipage}
\end{figure}

In multi-component BECs with more than 
four condensate wave functions, 
the number of possible molecules increases exponentially. 
Therefore, 
in order to classify such vortex molecules, 
we make use of the mathematical graph theory, 
in which vortices are expressed by vertices and 
the Rabi couplings are expressed by edges. 
Graphs isomorphic to each other are not distinguished 
in the graph theory. 
The number of independent connected graphs 
relevant for four-component BECs
is six. 
In the graph theory, a graph is characterized by the sequence of 
the number of edges connected to each vertex. 
For instance, $(1,1,2,2)$ implies that two vertices are connected by two edges, and other two vertices are connected by one edge.
We obtained numerical solutions 
for a four-component BEC with the same values of Rabi couplings 
(Fig.~\ref{fig:4comp}),
exhausting all possible graphs 
with four vortices as vertices 
including six connected graphs seen in Fig.~\ref{fig:4comp} 
and two disconnected graphs in Fig.~\ref{fig:4comp2}.  
The GP energy (\ref{eq:gp})
subtracted by the ground state energy for each configuration 
is shown at the top-right of 
each panel in Fig.~\ref{fig:4comp} and \ref{fig:4comp2}, 
where we chose 
$r=6$ for the spatial integration in Eq.~(\ref{eq:gp}). 

Our simulation indicate the occurrence of several new phenomena 
that do not exist in dimers or trimers:
1) twist, 2) holding, and 3) capture.
We have found that most molecules are accompanied by 
``twisted'' molecules, as in Fig.~\ref{fig:4comp} (a)--(e). 
A pair of untwisted and twisted molecules corresponds 
to identical graphs isomorphic to each other, 
while 
both pairs are energetically stable, corresponding to
absolute (left) and local (right) minima.
Once a twisted vortex molecule is formed, 
a certain amount of finite energy is required to ``untwist" it. 
The holding phenomenon and the absorption 
of a molecule inside 
a bigger molecule can be seen in Fig.~\ref{fig:4comp2} (a) and (b), 
respectively.

We can engineer as many multiple vortex molecules as we want with the constituent vortices. 
As an example, we presented vortex pentamers corresponding to all possible connected graphs with five vertices 
in a five-component BEC with the same Rabi couplings 
in Ref.~\cite{Eto:2013spa}. 

Thus far, we have concentrated on the case when 
$\omega_{ij}=\omega \neq 0$ and $\omega_{ij} = 0$  
for the connected and disconnected pair of the vortices,
respectively.
We can control the positions and shape of the molecule 
by varying $\omega_{ij}$ inhomogeneously. 
As an example, we present a vortex heptamer (seven vortices), 
designed as the orion in Fig.~\ref{fig:orion}.

\section{Multiple vortex molecules}
So far, we have discussed 
one integer vortex split into a molecule of fractional vortices. 
In this section, let us discuss multiple integer vortices. 
Vortices repel each other so that they can be stable 
under a rotation. 
We consider a rotating system $\Omega \neq 0$ with  
the trapping potential $V_{\rm ex} = m \omega_{\perp}^2 r^2$.
\subsection{Two components}
First, let us make comments on the cases 
without the Rabi couplings. 
For a few vortices in two-component BECs, 
vortex pairs constitute vortex polygons 
\cite{Kobayashi:2013wra}.
Vortex polygons have been found to be 
stable up to five molecules 
constituting a decagon   
and metastable for six molecules 
constituting a dodecagon.
However, seven or more molecules 
are unstable.
Polygons of vortex molecules have been also  
found in field theory \cite{Kobayashi:2013aja}.

For more than six molecules in two component BECs, 
the ground states are vortex lattice 
\cite{Mueller:2002,Kasamatsu:2003,Kasamatsu:2005,Aftalion:2011,
Aftalion:2012,Kuopanportti:2012}.
The phase diagram of the vortex lattice forming in the condensate was studied 
in Refs.~\cite{Mueller:2002,Kasamatsu:2003} 
and a rich variety of lattices was found.
The structure of the two component lattice depends on 
the sign and on the ratio $g_{12}/g \equiv \delta$ 
of the coupling constant of the inter/intra-component interactions. 
When $\delta <0$, the vortices of different components are attracted and are combined into integer vortices.
Because of repulsion among them, they organize in a triangular Abrikosov lattice. 
If $\delta>0$, depending on the value of $\delta$ and $\Omega$, vortices 
in each component organize in the triangular lattice, the square lattice, or the vortex-sheet.  
When $\delta=0$ the vortices of each component are organized in an Abrikosov lattice, but the vortices of different component are decoupled.
If $\delta$ is increased, the intercomponent interaction results in a repulsive force between the vortices of different components \cite{Eto:2011wp,Aftalion:2012} and the two component lattice has an hexagonal structure. 
When $\delta$ is increased further the unitary cell of the single component lattice is changed from a triangle to a square.
The value of $\delta$ for which the lattice reorganize depends on rotation speed.
If $\delta > 1$, phase separation occurs and vortex sheets appear \cite{Kasamatsu:2009}. 
Vortex lattices in two components with different masses were studied in Ref.~\cite{Kuopanportti:2012}.

\begin{figure*}
\begin{center}
\includegraphics[width=10cm]{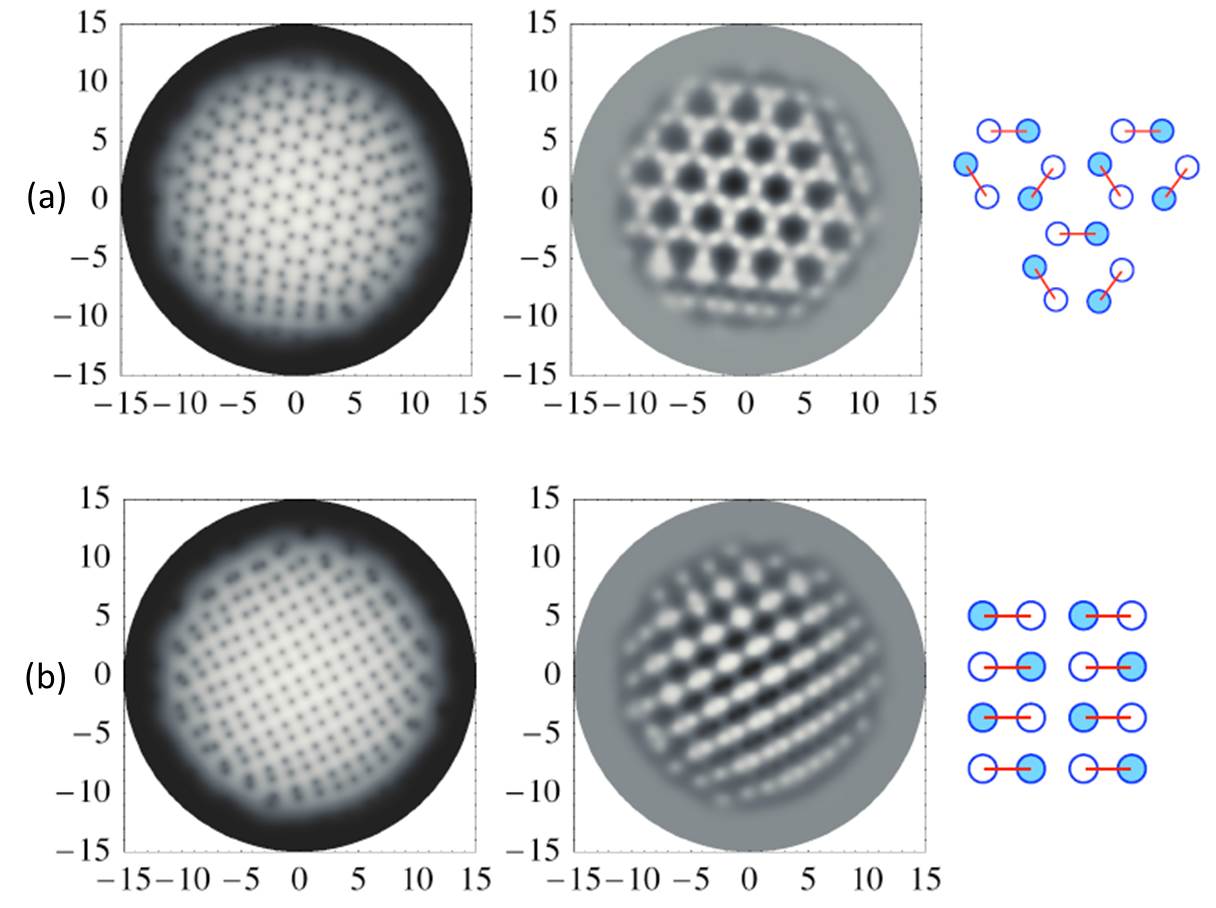}
\caption{Vortex lattices with a small Rabi coupling. 
(a) A triangular lattice and (b) a square lattice. 
The left panels are plots of the density profile of the condensate, 
$n=|\Psi_{1}|^{2}+|\Psi_{2}|^{2}$ (black dots are vortices 
in either of the two components), 
the middle panels are plots of the Rabi energy 
(white is positive, identified with domain walls), 
and  
the right panels are schematic drawings of the lattice structures,
where (un)shaded circles denote fractional vortices in first (second) components, 
and solid lines denote domain walls connecting 
two fractional vortices. 
\label{fig:molecule-lattice}} 
\includegraphics[width=12cm]{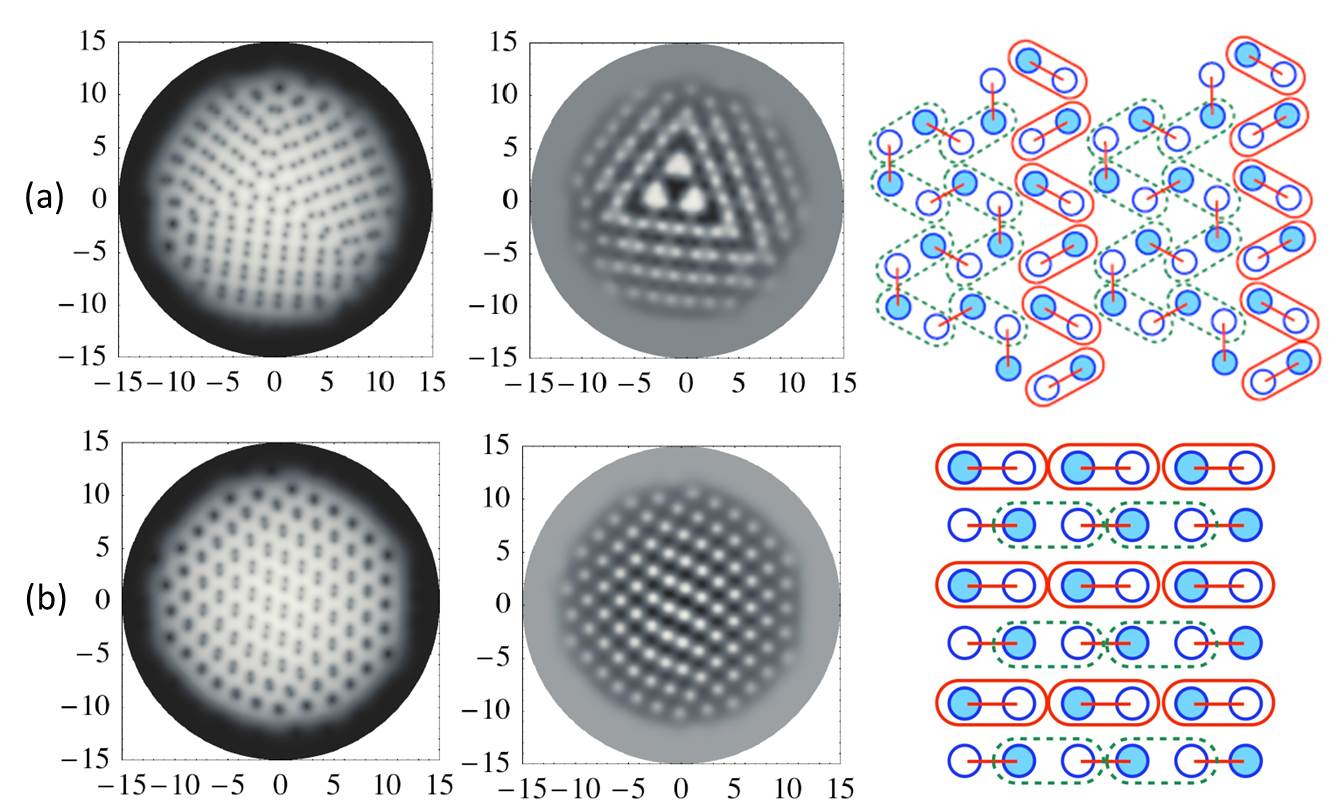}
\caption{Partner changing of vortices for (a) a triangular lattice 
and (b) a square lattice. 
Notations of the left and the middle panels are the same with 
Fig.~\ref{fig:molecule-lattice}.
In the right panels, 
the solid lines indicate the original molecules in 
Fig.~\ref{fig:molecule-lattice}, 
the solid round boxes do not change partners, and 
the dotted round boxes denote new partners. 
For the triangular lattice (a), there exist ${\mathbb Z}_3$ symmetric 
equivalent ways. In fact, all patterns appear in the simulation 
and there appear domain wall defects and a junction. 
\label{fig:partner-changing}}
\end{center}
\end{figure*}

So far, the Rabi couplings have not been included in most works 
on vortex lattices. 
A systematic study of the internal coherent (Rabi) coupling in vortex lattices
in two-component BECs under rotation 
has been given recently \cite{Cipriani:2013nya}.
In Fig.~\ref{fig:molecule-lattice}, we show typical configurations
for triangular and square lattices for 
small Rabi couplings. 
These vortex lattices must be transformed to 
the Abrikosov lattice of integer vortices at strong Rabi couplings. 
In the intermediate Rabi couplings,  
vortices change their partners 
in various ways depending on the inter-component coupling
to organize themselves for 
constituting the Abrikosov lattice of integer vortices. 
Typical examples of the partner changing for 
a triangular lattice and a square lattice 
are shown in Fig.~\ref{fig:partner-changing} (a) and (b), 
respectively. 
For the triangular lattice, there exist ${\mathbb Z}_3$ symmetric 
equivalent ways. 
In fact, all patterns appear in the simulation 
and there appear domain wall defects and a junction.

\subsection{Three components}

\begin{figure*}
\begin{center}
\begin{tabular}{cc}
\includegraphics[width=5.5cm]{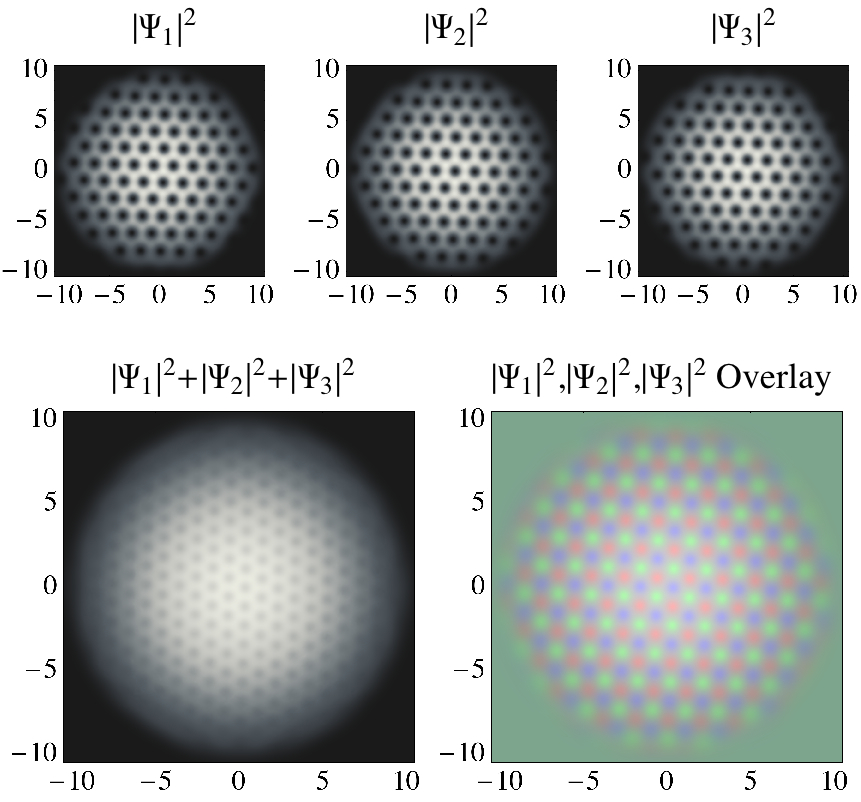} &
\includegraphics[width=5.5cm]{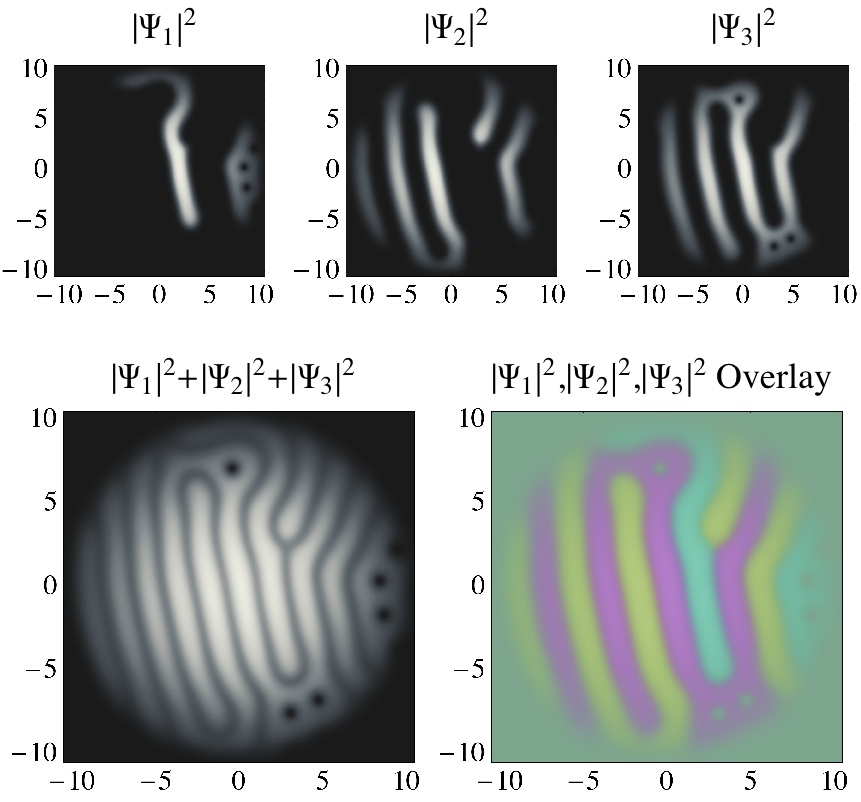} \\
(a) & (b)
\end{tabular} 
\caption{Vortex lattices (a) and sheets (b) in a three-component BEC. 
(a) The Abrikosov lattice of 1/3 quantized vortices for $\delta = 0.5$ (b) 
Vortex sheets for the phase separated region at $\delta =1.5$. 
}
\label{fig:3-comp-lattice}
\end{center}
\end{figure*}

Vortex lattices in a three-component BEC have been studied 
recently \cite{Cipriani:2013wia},  
where three kinds of fractional vortices winding one of three components are present. 
We have considered the symmetric case where 
all three intra-component (inter-component) couplings are the same 
$g_{11}=g_{22}=g_{33} \equiv g$ ($g_{12}=g_{23}=g_{31} \equiv \tilde g$) 
and all chemical potentials and
masses are the same 
as the case of  the mixture 
of atoms with different hyperfine states. 
Unlike the cases of two-component BECs 
where the phases of square and triangular lattices 
are present depending on the intercomponent 
coupling constant and the rotation speed, 
we find triangular ordered 
vortex lattices as in Fig.~\ref{fig:3-comp-lattice} (a), 
where 
three kind of fractional vortices are placed in 
order without defects, 
in all parameter region where 
the inter-component coupling $\tilde g$ is less 
than the intra-component coupling $g$. 

When $g>\tilde g$ on the other hand, 
we find the phase separation as in 
Fig.~\ref{fig:3-comp-lattice} (b). 
In a region where one component is present, 
the other two components must vanish, 
where we find ghost vortices in these two components 
whose positions are separated. 
We also find a domain wall junction. 

The introduction of the Rabi couplings remains as 
a future problem. 
However, since the vortex lattices found above are ordered  triangular lattices 
for $\tilde g < g$, 
they will be smoothly connected to the integer Abrikosov lattices
with increasing the Rabi couplings.


\section{Conclusion}
We have reported recent developments 
of vortex molecules in multi-component BECs. 
(1) Coherently coupled $N$-component BECs allow 
molecules of $N$ fractional vortices connected by 
sine-Gordon domain walls, representing graphs.  
(2) Vortex lattices in coherently coupled 
two-component BECs 
exhibit a vortex partner changing. 
(3) Abrikosov vortex lattices are robust 
in three-component BECs.

\begin{acknowledgements}
The work of M.~N.  is supported in part by 
a Grant-in-Aid for Scientific Research (
No. 25400268) 
and by the ``Topological Quantum Phenomena'' 
Grant-in-Aid for Scientific Research 
on Innovative Areas (
No. 25103720)  
from the Ministry of Education, Culture, Sports, Science and Technology 
(MEXT) of Japan. 
\end{acknowledgements}



\begin{thebibliography}{99}
\bibitem{Matthews}
M.~R.~Matthews, B.~P.~Anderson, P.~C.~Haljan, D.~S.~Hall,
C.~E.~Wieman, and E.~A.~Cornell, 
Phys.\ Rev.\ Lett..\ {\bf 83}, 2498 (1999).

\bibitem{Schweikhard:2004}
V.~Schweikhard, I.~Coddington, P.~Engels, S.~Tung, 
and E.~A.~Cornell
Phys.\ Rev.\ Lett.\ {\bf 93}, 210403 (2004). 

\bibitem{Ho:1998}
T.-L.~Ho, 
Phys.\ Rev.\ Lett.\ {\bf 81}, 742-745 (1998); 
T.~Ohmi and K.~Machida, 
J.\ Phys.\ Soc.\ Jpn.\ {\bf 67}, 1822 (1998);
  S.~Kobayashi, Y.~Kawaguchi, M.~Nitta and M.~Ueda,
Phys.\ Rev.\ A {\bf 86}, 023612 (2012); 
  J.~Lovegrove, M.~O.~Borgh and J.~Ruostekoski,
Phys.\ Rev.\ A {\bf 86}, 013613 (2012);
arXiv:1306.4700.

\bibitem{Semenoff:2006vv}
  G.~W.~Semenoff and F.~Zhou,
  Phys.\ Rev.\ Lett.\  {\bf 98} (2007) 100401;
%
  M.~Kobayashi, Y.~Kawaguchi, M.~Nitta and M.~Ueda,
  Phys.\ Rev.\ Lett.\  {\bf 103} (2009) 115301; 
  Y.~Kawaguchi, M.~Kobayashi, M.~Nitta and M.~Ueda,
Prog.\ Theor.\ Phys.\ Suppl.\  {\bf 186}, 455 (2010). 


\bibitem{Turner:2009}
A.~M.~Turner and E.~Demler, 
Phys.\ Rev.\ B {\bf  79}, 214522 (2009).


\bibitem{Mueller:2002}
E.~J.~Mueller and T.-L.~Ho,
Phys.\ Rev.\ Lett.\ {\bf 88}, 180403 (2002). 

\bibitem{Kasamatsu:2003}
K.~Kasamatsu, M.~Tsubota and M.~Ueda, 
Phys.\ Rev.\ Lett.\ {\bf 91}, 150406 (2003).

\bibitem{Kasamatsu:2004}
 K.~Kasamatsu, M.~Tsubota and M.~Ueda, 
Phys.\ Rev.\ Lett\ {\bf 93}, 250406 (2004).

\bibitem{Kasamatsu:2005}
	K.~Kasamatsu, M.~Tsubota and M.~Ueda, 
Int.\ J.\ Mod.\ Phys.\ {\bf B} 19, 1835 (2005).

\bibitem{Kasamatsu:2009}
K.~Kasamatsu, and M.~Tsubota,
Phys.\ Rev.\ {\bf A} 79, 023606 (2009).

\bibitem{Eto:2011wp} 
  M.~Eto, K.~Kasamatsu, M.~Nitta, H.~Takeuchi and M.~Tsubota,
Phys.\ Rev.\ A {\bf 83}, 063603 (2011).

\bibitem{Aftalion:2011}
P.~Mason and A.~Aftalion, 
Phys.\ Rev.\ A {\bf 84}, 033611 (2011). 

\bibitem{Aftalion:2012}
A.~Aftalion, P.~Mason and W.~Juncheng, 
Phys.\ Rev.\ A {\bf 85}, 033614 (2012).

\bibitem{Kuopanportti:2012}
P.~Kuopanportti, J.~A.~M.~Huhtam\"{a}ki, M.~M\"{o}tt\"{o}nen,
Phys.\ Rev.\ {\bf A} 85, 043613 (2012).

\bibitem{Eto:2012rc} 
  M.~Eto and M.~Nitta,
Phys.\ Rev.\ A {\bf 85}, 053645 (2012). 

\bibitem{Eto:2013spa} 
  M.~Eto and M.~Nitta,
arXiv:1303.6048 [cond-mat.quant-gas].  

\bibitem{Cipriani:2013nya} 
  M.~Cipriani and M.~Nitta,
arXiv:1303.2592 [cond-mat.quant-gas].  

\bibitem{Cipriani:2013wia} 
  M.~Cipriani and M.~Nitta,
Phys.\ Rev.\ {\bf A} (to appear)
[arXiv:1304.4375]


\bibitem{Tanaka:2001}
Y.~Tanaka, 
J.\ Phys.\ Soc.\ Jp.\ {\bf 70}, 2844 (2001);  
Phys.\ Rev.\ Lett.\ {\bf 88}, 017002 (2001).

\bibitem{Son:2001td}
  D.~T.~Son, M.~A.~Stephanov,
  Phys.\ Rev.\  A{\bf 65}, 063621 (2002).

\bibitem{Babaev:2002}
E.~Babaev, 
Phys.\ Rev.\ Lett.\ {\bf 89} (2002) 067001; 
E.~Babaev, A.~Sudbo and N.~W.~Ashcroft,
Nature {\bf 431}, 666 (2004); 
  J.~Smiseth, E.~Smorgrav, E.~Babaev and A.~Sudbo,
  Phys.\ Rev.\  B {\bf 71}, 214509 (2005); 
E.~Babaev and N.~W.~Ashcroft,
Nature Phys. {\bf 3}, 530 (2007).

\bibitem{Goryo:2007}
J.~Goryo, S.~Soma and H.~Matsukawa, 
 Euro Phys.\ Lett.\ {\bf 80}, 17002 (2007).

\bibitem{Nitta:2010yf} 
  M.~Nitta, M.~Eto, T.~Fujimori and K.~Ohashi,
J.\ Phys.\ Soc.\ Jap.\  {\bf 81}, 084711 (2012).

\bibitem{Kobayashi:2013wra} 
  M.~Kobayashi and M.~Nitta,
arXiv:1307.1345 [cond-mat.quant-gas].  



\bibitem{Kobayashi:2013aja} 
  J.~Jaykka and M.~Speight,
Phys.\ Rev.\ D {\bf 82}, 125030 (2010); 
  M.~Kobayashi and M.~Nitta,
Phys.\  Rev.\ D {\bf 87}, 125013 (2013).

\end{thebibliography}
\end{document}